\documentclass[prl,reprint,amsmath,amssymb,aps,preprintnumbers,showpacs]{revtex4-1}
\usepackage[colorlinks=true,linkcolor=black, citecolor=black,urlcolor=black]{hyperref} 
\usepackage{amstext,amsmath,amssymb,amsfonts,mathrsfs}   
\usepackage{graphicx}  
\usepackage{braket} 
\usepackage{bbm} 
\usepackage[british,UKenglish,USenglish,english,american]{babel}

\newcommand{\AdS}{\text{AdS}}
\renewcommand{\S}{\text{S}}
\newcommand{\T}{\text{T}}

\newcommand{\alg}[1]{\text{#1}}

\newcommand{\so}{\alg{so}}
\renewcommand{\sl}{\alg{sl}}
\renewcommand{\u}{\alg{u}}
\newcommand{\su}{\alg{su}}

\newcommand{\de}{\text{d}}

\newcommand{\eg}{\textit{e.g.}}

\newcommand{\Ham}{\mathbf{H}}
\newcommand{\action}{\mathbf{S}}

\newcommand{\param}{s}

\begin{document}

\vspace*{0cm}

\title{Strings on NS-NS Backgrounds as Integrable Deformations}
\author{Marco Baggio$^{1}$}
\email{marco.baggio@kuleuven.be}
\author{Alessandro Sfondrini$^{2}$}
\email{sfondria@itp.phys.ethz.ch}

\affiliation{$^1$ Instituut voor Theoretische Fysica, KU Leuven, Celestijnenlaan 200D, B-3001 Leuven, Belgium \\
${}^2$ Institut f\"ur theoretische Physik, ETH Z\"urich, Wolfgang-Pauli-Stra{\ss}e 27, 8093 Z\"urich,
Switzerland}

\begin{abstract}
We consider the worldsheet S~matrix of superstrings on an $\AdS_3\times\S^3\times\T^4$ NS-NS background in uniform light-cone gauge. We argue that scattering is given by a CDD factor that is non-trivial only between opposite-chirality particles, yielding a spin-chain-like Bethe Ansatz. Our proposal reproduces the spectrum of non-protected states computed from the Wess-Zumino-Witten description and the perturbative tree-level S~matrix. This suggests that the model is an integrable deformation of a free theory similar to those arising from the $T\bar{T}$ composite operator.
\end{abstract}


\maketitle

\paragraph{Introduction.}
The study of the AdS/CFT correspondence~\cite{Maldacena:1997re, Witten:1998qj,Gubser:1998bc} gave new energy to the search for exactly solvable string backgrounds. For a general string background only few observables may be computed exactly, usually owing to supersymmetry. Notable exceptions are plane-wave backgrounds~\cite{Blau:2001ne,Blau:2002dy,Berenstein:2002jq} where the light-cone Hamiltonian is free, Wess-Zumino-Witten (WZW) models~\cite{Witten:1983ar,Knizhnik:1984nr,Gepner:1986wi,Maldacena:2000hw} where current algebras can be used to solve the theory, and integrable backgrounds~\cite{Arutyunov:2009ga,Beisert:2010jr,Sfondrini:2014via} where the worldsheet scattering in light-cone gauge factorises along the lines of Ref.~\cite{Zamolodchikov:1978xm}.
The best-understood integrable AdS background is $\AdS_5\times\S^5$ supported by Ramond-Ramond (R-R) five-form fluxes. The string equations of motion are integrable~\cite{Bena:2003wd} and the factorised S~matrix can be computed from symmetry considerations~\cite{Beisert:2005tm,Beisert:2006ez,Arutyunov:2006yd}. The spectrum follows from imposing periodic boundary conditions and accounting for finite-size ``wrapping'' effects~\cite{Ambjorn:2005wa}, leading eventually to a quantum spectral curve~\cite{Gromov:2013pga}. Remarkably, the integrability approach to $\AdS_5\times\S^5$ superstrings  can be extended to three-~\cite{Basso:2015zoa} and higher-point~\cite{Eden:2016xvg,Fleury:2016ykk} correlation functions, and even to non-planar corrections~\cite{Eden:2017ozn,Bargheer:2017nne}, though the treatment of wrapping corrections is less thoroughly understood in that context.
Another important class of integrable backgrounds is given by $\AdS_3\times\S^3\times\T^4$ and $\AdS_3\times\S^3\times\S^3\times\S^1$ geometries supported by R-R and Neveu-Schwarz-Neveu-Schwarz (NS-NS) three-form fluxes. They are also classically integrable~\cite{Babichenko:2009dk,Cagnazzo:2012se} and their S~matrix can be fixed by symmetries~\cite{Borsato:2012ud,Borsato:2013qpa, Borsato:2014exa}, even for mixed background fluxes~\cite{Lloyd:2014bsa,Borsato:2015mma}. The purely R-R backgrounds resemble $\AdS_5\times\S^5$: the S~matrix has a complicated scalar factor~\cite{Borsato:2013hoa,Borsato:2016xns} and the dispersion relation is periodic~\cite{Borsato:2012ud} suggesting a dual spin-chain interpretation similar to Ref.~\cite{Minahan:2002ve}. Instead the NS-NS flux yields a \emph{linear} contribution to the dispersion~\cite{Hoare:2013lja,Lloyd:2014bsa}. The pure-NS-NS model corresponds to a supersymmetric WZW model and the S~matrix should simplify drastically there. The analysis of light-cone gauge symmetries of Ref.~\cite{Lloyd:2014bsa}, valid for generic mixtures of R-R and NS-NS fluxes, is insufficient to determine the S~matrix at the WZW point. In this letter we analyse the $\AdS_3\times\S^3\times\T^4$ WZW model in uniform light-cone gauge~\cite{Arutyunov:2004yx,Arutyunov:2005hd,Arutyunov:2006gs}, considering its classical bosonic Hamiltonian, its spectrum and its S~matrix. We observe that shifting the gauge parameter has a similar effect to a $T\bar{T}$ deformation~\cite{Zamolodchikov:2004ce,Smirnov:2016lqw,Cavaglia:2016oda} and that the $\T^4$ sector of the theory is free in a suitable gauge. This motivates us to further investigate the spectrum to determine if it can be related to that of a free theory. 
We find that in the ``spectrally unflowed'' sector~\cite{Maldacena:2000hw} the energies of non-protected states can be reproduced from the Bethe-Yang equations by adding a CDD factor~\cite{Castillejo:1955ed} to a free theory. The resulting S~matrix coincides at tree level with the known perturbative result~\cite{Hoare:2013pma} in a suitable gauge. Moreover we argue that, due to supersymmetry, wrapping corrections cancel out similarly to what happened for protected states in Ref.~\cite{Baggio:2017kza}, so that the Bethe-Yang equations are exact. The CDD factor appearing in our construction is exactly that of a $T\bar{T}$ deformation when we restrict to $\T^4$~\footnote{%
Such a CDD factor first appeared in Ref.~[39] in the context of uniform light-cone gauge transformations.
}; in general however it differs from it due to the presence of an additional $\u(1)$ current. The simple form of the S~matrix makes the study of this NS-NS background almost as straightforward as that of a plane-wave one, paving the way to a wealth of explicit computations.
We conclude this letter by detailing additional checks of our proposal, which we intend to present in an upcoming publication~\cite{companion}, as well as by commenting on several possible future directions.

\paragraph{WZW model in light-cone gauge.}
The Green-Schwarz action for $\AdS_3\times\S^3\times\T^4$ with mixed NS-NS and R-R three-form fluxes has been analysed in some detail in Ref.~\cite{Lloyd:2014bsa}. We restrict to the pure NS-NS case corresponding to the WZW model, and find the bosonic action
\begin{equation}
\action =-\frac{k}{2}\int\limits_{-\infty}^{+\infty}\!\de\tau\! \int\limits_{0}^{R}\!\de\sigma \big(
\gamma^{\alpha\beta}G_{\mu\nu}+\epsilon^{\alpha\beta}B_{\mu\nu}\big)
\partial_\alpha X^\mu\partial_\beta X^\nu,
\end{equation}
where $k$ is the WZW level, $\gamma^{\alpha\beta}$ is the worldsheet metric with $|\gamma|=-1$, $G_{\mu\nu}$ is the $\AdS_3\times\S^3\times\T^4$ metric and $B_{\mu\nu}$ is the Kalb-Ramond field. We write the line element as
\begin{eqnarray}
\label{eq:metric}
\de s^2&=&
\,-(1+|z|^2)\de t^2+(1-|y|^2)\de\phi^2+\de x^j \de x^j\\
&&\!\! +\big(\delta_{ij}-\frac{z_iz_j}{1+|z|^2}\big)\de z^i\de z^j
+\big(\delta_{ij}+\frac{y_iy_j}{1-|y|^2}\big)\de y^i\de y^j,
\nonumber
\end{eqnarray}
where $t,z_1,z_2$ are in $\AdS_3$, $\phi,y^3,y^4$ describe $S^3$ and $x^5,\dots x^8$ give~$\T^4$. The Kalb-Ramond field is given by $B=\epsilon^{ij}z_i dz_j\wedge dt- \epsilon^{ij}y_i dy_j\wedge d\phi$. Fermions couple to $H=dB$, see Refs.~\cite{Cvetic:1999zs,Wulff:2013kga,Lloyd:2014bsa} for explicit formulae. To fix uniform light-cone gauge~\cite{Arutyunov:2004yx,Arutyunov:2005hd,Arutyunov:2006gs} we introduce, for $0\leq a \leq 1$,
\begin{equation}
x^+=(1-a)t+a\phi, \qquad x^-=\phi-t\,,
\end{equation}
and following \eg~Ref.~\cite{Arutyunov:2009ga} we introduce conjugate momenta $p_\mu=\delta\action/\delta(\partial_0 X^\mu)$ and fix
\begin{equation}
x^+ = \tau,\qquad p_-=(1-a)p_\phi-ap_t=k\,.
\end{equation}
This breaks conformal invariance and in particular fixes the worldsheet size~$R$
\begin{equation}
\label{eq:Rdep}	
R = (1-a)\int\limits_{0}^{R} \de\sigma p_\phi - a\int\limits_{0}^{R} \de\sigma p_t= J +a(E-J)\,,
\end{equation}
where $E$ is the string energy and $J$ its angular momentum. The light-cone Hamiltonian is
\begin{equation}
\label{eq:lcHam}
\Ham = -\int\limits_{0}^{R} \de\sigma\,p_+ = E-J\,,
\end{equation}
and the $\AdS_3\times\S^3$ BPS bound guarantees $\Ham\geq0$.
Notice that $-p_+$ is $a$-dependent, and gauge invariance dictates
\begin{equation}
\label{eq:gaugeinv}
\frac{\de}{\de a}\int\limits_{0}^{R(a)} \de\sigma\, p_+(a)  = 0\,.
\end{equation}
The density $p_+(a)$ can be easily found as in Ref.~\cite{Lloyd:2014bsa} by solving the Virasoro constraints. Truncating it to $\T^4$ modes and setting $\param=(a-1/2)/k$ we find
\begin{equation}
\label{eq:flatHamiltonian}
\Ham\big|_{\T^4}=\int\limits_0^{R(\param)}\de\sigma\,
\frac{
1-\sqrt{1-4\param\, H_{\text{free}}+4\param^2\,(p_j\acute{x}^j)^2}}{2\,\param},
\end{equation}
with $H_{\text{free}}=(p_jp^j+k^2\acute{x}^j\acute{x}_j)/2k$. Eqs.~\eqref{eq:flatHamiltonian} reduces to a free Hamiltonian at $\param=0$, \textit{i.e.}\ at~$a=1/2$. In view of Eqs.~(\ref{eq:Rdep}--\ref{eq:flatHamiltonian}) we conclude that the $\T^4$ modes can be equivalently represented as a free system with state-dependent worldsheet length $R=J+\Ham/2$ (for $a=1/2$) or as an interacting one with fixed length $R=J$ (for $a=0$).

\paragraph{Relation to $T\bar{T}$ deformations.}
The form of Eq.~\eqref{eq:flatHamiltonian} is that of a $T\bar{T}$ deformation of free bosons~\cite{Zamolodchikov:2004ce,Smirnov:2016lqw,Cavaglia:2016oda}. 
To understand why, let us review and in fact slightly generalise the construction of such deformations. Given two conserved local currents $j_I^\alpha$, $I=1,2$ the limit
\begin{equation}
\label{eq:jjdeform}
j_1j_2(x)=\lim_{y\to x} j_1^\alpha(x)\,j_2^\beta(y)\,\epsilon_{\alpha\beta}\,,
\end{equation}
is well-defined owing to the arguments of Ref.~\cite{Zamolodchikov:2004ce}, and 
\begin{equation}
\langle j_1j_2 \rangle=\langle j_1^\alpha\rangle\,\langle j_2^\beta\rangle \,\epsilon_{\alpha\beta}\,.
\end{equation}
Notice that we do not require any of the currents $j_I^\alpha$ to be chiral.
A $T\bar{T}$ deformation corresponds to the case $j_{I}^\alpha=T^{\alpha I}$. Coupling~$T^{\alpha I}$ to a $\u(1)$ current yields deformations of the type considered in Ref.~\cite{Guica:2017lia}; ``$J\bar{J}$'' deformations fall in this class too, by taking a current and its (conserved) Hodge dual. For such special choices of $j_{I}^\alpha$ the deformation has a simple effect on the spectrum~\cite{Zamolodchikov:2004ce,Guica:2017lia}, and in particular for a $T\bar{T}$ deformation of parameter~$\alpha$ we have
\begin{equation}
\label{eq:TTbareq}
\partial_{\alpha} H_n = -H_n \partial_R H_n\,,
\end{equation} 
for a state $|n\rangle$ of energy $H_n$ and zero momentum~\cite{Zamolodchikov:2004ce}. From this it follows that $H_n(R,\alpha)=H_n(R-\alpha H_n,0)$: the deformation amounts to a state-dependent shift of the length, which can be described as a CDD factor~\cite{Smirnov:2016lqw,Cavaglia:2016oda}. This is also the effect induced on $p_+$ by $a$-gauge transformations, \textit{cf.}~Eq.~\eqref{eq:gaugeinv}, which explains the form of Eq.~\eqref{eq:flatHamiltonian}. Gauge transformations and $T\bar{T}$ deformations should not be confused however: the former \emph{leave the spectrum invariant} while the latter correspond to changing $p_+$ while leaving $R$ fixed or viceversa. Indeed the differential equation for $a$-gauge transformations is Eq.~\eqref{eq:gaugeinv} rather than Eq.~\eqref{eq:TTbareq}.
Hence our observation that the ``flat'' subsector of classical $\AdS_3\times\S^3\times\T^4$ strings is simply related to a free theory is unsurprising in view of Refs.~\cite{Smirnov:2016lqw,Cavaglia:2016oda}. It is remarkable, as we will see, that this extends to the full quantum level, not only for $\T^4$ modes but for the whole superstring background.

\paragraph{WZW spectrum.}
The spectrum of the light-cone Hamiltonian can be constructed from the left and right Ka\v{c}-Moody currents~\cite{Giveon:1998ns,Maldacena:2000hw, Pakman:2003cu,Israel:2003ry, Raju:2007uj,Ferreira:2017pgt}, as we very briefly review here. In the left sector we have $\sl(2)_{k+2}$ currents $L^{0,\pm}_{-n}$, $\su(2)_{k-2}$ currents $J^{3,\mp}_{-n}$, torus modes $\alpha^r_{-n}$ as well as their fermionic superpartners. They act on a vacuum $|\ell_0,j_0\rangle$ given by a highest (resp.~lowest) weight state of $\sl(2)$, resp.~$\su(2)$. A generic state is obtained by acting with positive energy modes of the currents ($n\geq0$) on the vacuum, \textit{e.g.}
\begin{equation}
\prod_{i=1}^{\ell^{+}}\!L^{+}_{-n_i}
\prod_{i=1}^{\ell^{-}}\!L^{-}_{-n_i}
\prod_{i=1}^{ j^{+}}\!J^{+}_{-n_i}
\prod_{i=1}^{j^{-}}\!J^{-}_{-n_i}
\prod_{i=1}^{M_T}\!\alpha^{r_i}_{-n_i}
|\ell_0,j_0\rangle.
\end{equation}
Such a state has (left) energy $\ell=\ell_0+\delta \ell$ and (left) angular momentum $j=j_0 - \delta j$ with $\delta\ell=\ell^+-\ell^-$ and $\delta j=j^--j^+$. Fermions can be added in a similar way and the usual subtleties arise depending on their boundary conditions, see \textit{e.g.}~Refs.~\cite{Ferreira:2017pgt,upcomingAndrea} for details. Physical states are subject to restrictions, the most important being the \emph{mass-shell condition}
\begin{equation}
\label{eq:massshell}
-\frac{\ell_0(\ell_0-1)}{k}+\frac{j_0(j_0+1)}{k}+N_{\text{eff}}=0\,,
\end{equation}
where $N_{\text{eff}}=\sum_j n_j$ is the total mode number (which in the NS sector is shifted by $-1/2$).
More general sectors of the spectrum can be described by spectral flow~\cite{Maldacena:2000hw}, though we will not consider them in this letter.
 Similar expressions hold in the right sector, which we label with tildes. Imposing level-matching $N_{\text{eff}}=\tilde{N}_{\text{eff}}$ and $j_0=\tilde{\jmath}_0$ we finally get
\begin{equation}
\label{eq:WZWlcenergy}
E-J =\sqrt{(2j_0+1)^2+4kN_{\text{eff}}}-(2j_0+1)+\delta\,,
\end{equation}
where $\delta=\delta\ell+\delta\tilde{\ell}+\delta j+\delta\tilde{\jmath}+2$ and we solved Eq.~\eqref{eq:massshell}. Notice that for BPS states $E=J$ and $N_{\text{eff}}=\delta=0$
\footnote{This comes about slightly differently in the R and NS sectors and requires care with the GSO projection.}.

\paragraph{Free-theory interpretation}
The spectrum of excitations over the BPS ``vacuum'' is simply related to a $T\bar{T}$-like deformation of a free theory. Consider a theory of eight free bosons with dispersion
\begin{equation}
\label{eq:dispersion}
E(p) = \Big|\frac{k}{2\pi} p + \mu\Big|\,,\qquad
\mu\in\{0,0,+1,-1\}^{\oplus 2}\,.
\end{equation}
This coincides 
\footnote{The four modes with $\mu=0$ correspond to $\T^4$ directions; introducing complex coordinates $z,\bar{z},y,\bar{y}$ for the $\AdS_3\times S^3$ transverse fields, two fields $(z,y)$ have $\mu=1$ and their conjugate have $\mu=-1$.}
 with the plane-wave dispersion of our string background~\cite{Berenstein:2002jq, Russo:2002rq,upcomingAndrea} which for the NS-NS background is \emph{exact}~\cite{Hoare:2013lja,Lloyd:2014bsa}. Supersymmetry can be realised by adding eight fermions with the same masses~$\mu$. Imposing boundary conditions on a circle of size~$R$ we have
\begin{equation}
\label{eq:totalH}
\Ham=\sum_{i} E\Big(\frac{2\pi n_i}{R}\Big)
=\frac{k (N+\tilde{N})}{R} + \sum_{i}\mu_i\, \text{sgn}(n_i)\,,
\end{equation}
where we split left- and right-movers. Notice that to remove the absolute value we have assumed that $R\leq k$; we will see shortly why this is the case.
If we now postulate the \textit{state-dependent length}
\begin{equation}
\label{eq:Ldependence}
R = R_0 + \frac{\Ham-\boldsymbol{\mu}}{2}\,, \qquad \boldsymbol{\mu}=\sum_{i}\mu_i\, \text{sgn}(n_i)\,,
\end{equation}
and solve Eq.~\eqref{eq:totalH} we precisely reproduce the WZW light-cone energy~\eqref{eq:WZWlcenergy} with the following identifications. Firstly, $\Ham=E-J$ as in Eq.~\eqref{eq:lcHam}.
Next $R_0=2j_0+1$ is the $J$-charge of the BPS state in the R-R sector corresponding to the middle of the $\T^4$ Hodge diamond. Notice that taking $R_0$ to be the charge of a reference vacuum rather than the $J$-charge of the state itself mimics the dual spin-chain construction for~$\AdS_5\times\S^5$~\cite{Minahan:2002ve,Beisert:2005fw}.
Finally $\boldsymbol{\mu}=\delta$. Let us justify this. Notice that when no excitations on $\AdS_3\times\S^3$ are present, $\boldsymbol{\mu}=0$ and Eq.~\eqref{eq:Ldependence} precisely describes a $T\bar{T}$ deformation~\cite{Zamolodchikov:2004ce,Smirnov:2016lqw,Cavaglia:2016oda}. Consider now a state with some $\T^4$ excitations over the BPS vacuum and a single $\S^3$ mode, say $J^{\pm}_{-n}$. For the charges to match this should correspond to a boson with $p=2\pi n/R\geq 0$ and $\mu=\mp 1$; conversely $\tilde{J}^{\pm}_{-\tilde{n}}$ gives a boson with $p=-2\pi\tilde{n}/R\leq 0$ and $\mu=\pm1$
\footnote{Zero-modes require some care. In the WZW model $J^+_0,\tilde{J}^+_0$ annihilate the vacuum; correspondingly, for $\mu=1$ (resp.\  $\mu=-1$) only the left-moving (resp.\ right-moving) zero mode survive.}.
This matches the identification of~$\mu$ with the $\su(2)$-spin of $\S^3$ excitations in the plane-wave limit~\cite{Lloyd:2014bsa}.
The other bosons as well as the fermions can be similarly described and will be presented elsewhere~\cite{companion}.
Finally, notice that $R=\ell_0+j_0$ with our identifications. The condition $R\leq k$ which we used to remove the absolute values in Eq.~\eqref{eq:totalH} follows from the unitarity bounds of the WZW model~\cite{Maldacena:2000hw}. Sectors with larger values of~$R$ should arise from spectral flow, see also Ref.~\cite{upcomingAndrea} for a discussion of this fact in the plane-wave limit.

\paragraph{S matrix and Bethe-Yang equations.}
An energy-dependent shift of the length can be described as a CDD factor~\cite{Castillejo:1955ed} to the S~matrix, see Ref.~\cite{Smirnov:2016lqw}. This is also the case for the shift of Eq.~\eqref{eq:Ldependence} which corresponds to a CDD factor whose phase is
\begin{equation}
\Phi_{jk}=
p_j\,E_k-p_k\,E_j
-p_j\,\hat{\mu}_k
+p_k\,\hat{\mu}_j\,,
\end{equation}
where $\hat{\mu}=\mu\,\text{sgn}(kp+2\pi\mu)$.
Starting from a free theory, we get a \emph{diagonal} S~matrix with elements $S_{jk}=\exp(\tfrac{i}{2}\Phi_{jk})$. The Bethe-Yang equations follow immediately,
\begin{equation}
\label{eq:BetheYang}
1=\exp(i p_k R_0)\prod_{j\neq k}S_{kj}=\exp(i p_k R)\,.
\end{equation}
where in the last equation we used the level-matching condition $\sum p_j=0$. Given that $E$ and $\hat{\mu}$ distinguish between left- and right-movers, it is convenient to treat such modes separately. We introduce labels ``$\pm$'' for particles having $\partial E/\partial p =\pm k/2\pi$, yielding four cases for the S~matrix. We get
\begin{equation}
\label{eq:Smatrix}
S_{jk}^{++}=S_{jk}^{--}=1,\quad
S_{jk}^{-+}=\exp \left(i\tfrac{k}{2\pi}p_j p_k\right)=\frac{1}{S_{kj}^{+-}}\,.
\end{equation}
This illustrates the role of $\boldsymbol{\mu}$: it makes the left-left and right-right scattering trivial, as we would expect in a theory where particles move at the speed of light. Notice that such scattering is much simpler than the one arising in Refs.~\cite{Zamolodchikov:1992zr,Fendley:1993xa}, where non-diagonal and non-perturbative  left-left and right-right S~matrices appear.
 These expressions match the perturbative tree-level result for $S^{\pm\mp}_{ij}$ of Ref.~\cite{Hoare:2013pma}. To compare our expressions we should firstly take the results of Ref.~\cite{Hoare:2013pma} in the  $a=0$ gauge; in that case, the length in the Bethe-Yang equations is the $J$-charge of the state---in contrast to our conventions, in which it is the $J$-charge \textit{of the BPS vacuum}. Accounting for these different conventions is akin to going from the string-frame to the spin-chain frame~\cite{Arutyunov:2006yd, Borsato:2012ud,Borsato:2013qpa}. With these identifications, the left-right and right-left S~matrices match with Ref.~\cite{Hoare:2013pma}
 \footnote{\textit{Ibidem} $S^{\pm\pm}_{ij}$ is seemingly non-trivial, at least for mixed-flux backgrounds; however, carefully taking the pure-NS-NS limit as explained around Eq.~(3.24) there, one finds $S^{\pm\pm}_{ij}=1$, coherently with the expectation from perturbation theory.}.
Based on the integrability treatment of strings in flat space~\cite{Dubovsky:2012wk} it may appear surprising that our analysis relies solely on the Bethe-Yang equations~\eqref{eq:BetheYang} and does not require the mirror thermodynamic Bethe Ansatz to account for finite-size effect, \textit{cf.}\ Ref.~\cite{Ambjorn:2005wa,Arutyunov:2007tc}; this is all the more concerning given that this background features gapless excitations that usually lead to severe wrapping effects~\cite{Abbott:2015pps}. This simplification is due to supersymmetry: as the scattering is diagonal, wrapping corrections~\cite{Luscher:1985dn, Luscher:1986pf, Bajnok:2008bm} to a state with momenta $p_1,\dots p_M$ take a simple form
\begin{equation}
\label{eq:wrapping}
\int \de\rho\,e^{-\varepsilon(\rho)L} \sum_{X} (-1)^{F_X} \prod_{j=1}^MS_{Xj}(\rho,p_j)\,,
\end{equation}
where $X$ is any virtual particle.
Regardless of the details, here bosons and fermions come in pairs with identical dispersion and scattering, so that \emph{the integrand vanishes}; this is the same argument that guarantees that BPS states are immune from wrapping corrections in Ref.~\cite{Baggio:2017kza}.

\paragraph{Towards a deformation of the full action.}
A formula for the action of $T\bar{T}$ deformations of scalar field theories is known~\cite{Cavaglia:2016oda,Tateo:talk,upcomingTateo}. We briefly discuss two subtleties arising when applying such an approach here: firstly, our transformation involves the current $\boldsymbol{\mu}$; secondly, our free action has the $\mu$-dependent dispersion~\eqref{eq:dispersion}.
Na\"ively we would use Eq.~\eqref{eq:jjdeform} with one of the currents given by~$j^\alpha$ such that $\int \de\sigma j^0=\boldsymbol{\mu}$; unfortunately, while such a conserved current exists in a free theory, it is non-local and Zamolodchikov's arguments~\cite{Zamolodchikov:2004ce} do not apply
\footnote{Additionally here extended supersymmetry results in a degenerate spectrum, which also prevents us from straightforwardly applying Zamolodchikov's arguments.}. 
Alternatively we can ask whether the gauge-fixed WZW action is the $T\bar{T}$ deformation of \textit{some} simpler theory; this is also quite subtle. In the presence of several $\so(2)$ symmetries such as the ones rotating $z_{1,2}$ and $y_{3,4}$ the stress-energy tensor is not uniquely defined. To be concrete, we truncate our theory to the $S^3$ modes and introduce complex coordinates $y,\bar{y}$. 
The dispersion~\eqref{eq:dispersion} can be reproduced by coupling $y,\bar{y}$ to a constant $\u(1)$ background gauge field~$A^\alpha$. The Noether stress-energy tensor~$T^{\alpha\beta}_{\text{N}}$ \emph{is not gauge-invariant}; adding improvement terms yields the Hilbert stress-energy tensor~$T^{\alpha\beta}_{\text{H}}$. The difference of the two $T\bar{T}$ operators is also of the form~\eqref{eq:jjdeform},
\begin{equation}
T_{\text{H}}\bar{T}_{\text{H}}-T_{\text{N}}\bar{T}_{\text{N}}=
\epsilon_{\alpha\beta}\epsilon_{IJ}j^{\alpha;I}T^{\beta J}\,,
\end{equation}
where the two currents $j^{\alpha;I}$ are related to the components of the constant gauge field
\begin{equation}
j^{\alpha;I}= i A^I\, (p^\alpha \bar{y}- \bar{p}^\alpha y)\,.
\end{equation}
We hence have at least two \textit{inequivalent} $T\bar{T}$ deformations. \textit{A priori} it is unclear which one is more natural; interestingly, a Hilbert-$T\bar{T}$ deformation relates the gauge-fixed GS action to a simple sigma~model action for the sphere fields
\footnote{Here the Hilbert-$T\bar{T}$ deformation leads to formulae similar to Cavagli\`a \textit{et al.}~with $\partial_\alpha\to D_\alpha$.
}
\begin{equation}
\label{eq:trucatedaction}
\action\big|_{\S^3}=k\int\limits_{-\infty}^{+\infty}\de\tau \int\limits_0^R \de\sigma\,\eta^{\alpha\beta} \frac{D_\alpha y\,D_\beta \bar{y}}{1-y\bar{y}}\,,
\end{equation}
with background gauge field~$A^\alpha = g^{-1}\partial^\alpha g$, $g=\exp[i  \sigma]$. The integrability of the classical $\AdS_3\times\S^3\times\T^4$ action suggests that~\eqref{eq:trucatedaction} is classically integrable~too.

\paragraph{Conclusions and outlook.}
We have found evidence that superstrings on $\AdS_3\times\S^3\times\T^4$ with NS-NS three-form flux are described by a simple integrable theory of eight relativistic bosons and fermions with dispersion~\eqref{eq:dispersion} and S~matrix~\eqref{eq:Smatrix} given by a CDD factor. For such a theory wrapping corrections cancel and the Bethe-Yang equations are exact, rather than asymptotic. While this description is strongly reminiscent of a $T\bar{T}$ deformation, constructing the appropriate perturbing operator is quite subtle.
A number of questions immediately arise. Our construction here was limited to ``unflowed'' sector of the WZW model, corresponding to $R\leq k$ in Eq.~\eqref{eq:totalH}. It would be interesting to extend this to the $w$-th spectrally flowed sector corresponding to $wk<R\leq (w+1)k$, see also Ref.~\cite{upcomingAndrea} for a discussion of this in the plane-wave limit; notice that when the inequality is saturated the mass-gap in Eq.~\eqref{eq:totalH} vanishes and new gapless modes appear. We also restricted to states with vanishing total momentum (\textit{i.e.}~$N=\tilde{N}$). In light-cone gauge, winding sectors should also be included~\cite{Arutyunov:2004yx,Arutyunov:2009ga}, which would modify our analysis and in particular Eq.~\eqref{eq:BetheYang}. Finally it is intriguing that the gauge-fixed WZW action is related to Eq.~\eqref{eq:trucatedaction} and it would be worth exploring more such a sigma~model. We will return to these questions in an upcoming publication~\cite{companion}. It would also be worth extending this analysis to $\AdS_3\times\S^3\times\S^3\times\S^1$ backgrounds, whose integrability~\cite{Cagnazzo:2012se,Borsato:2015mma,Baggio:2017kza} and WZW~\cite{Maldacena:2000hw,Eberhardt:2017fsi} descriptions are well-established, as well as more general supersymmetric theories with diagonal scattering, where wrapping corrections are also expected to cancel. It looks less likely that this scenario might hold for mixed R-R and NS-NS backgrounds, as the S~matrix is non-trivial in that case~\cite{Hoare:2013pma,Lloyd:2014bsa}, though one might hope that the first correction in the R-R flux is captured by Eqs.~(\ref{eq:totalH}--\ref{eq:Ldependence}) with the exact mixed-flux dispersion~\cite{Hoare:2013lja,Lloyd:2014bsa} instead of Eq.~\eqref{eq:dispersion}. Such mixed-flux dynamics is particularly interesting as it captures a large part of the moduli space~\cite{upcomingBogdan}.
Describing strings on NS-NS backgrounds as simple integrable theories would have a number of interesting applications. As our description depends parametrically on the WZW level~$k$ we could apply it to \textit{e.g.}~the semi-classical limit $k\gg1$ as well as to special cases such as the $k=1$ theory which was recently related to a symmetric-product orbifold CFT~\cite{Giribet:2018ada, Gaberdiel:2018rqv}. Interestingly, our dispersion~\eqref{eq:dispersion} at $k=1$ precisely describes the single-excitation spectrum of the symmetric-product orbifold CFT of $\T^4$~\cite{Sfondrini:talk}. This might help us find an integrability description for symmetric-product orbifold~CFTs, \textit{cf.}\ also Ref.~\cite{Pakman:2009zz}.
It would also be interesting to extend this map beyond the spectrum: recently integrability techniques have been developed to compute three-~\cite{Basso:2015zoa} and higher-point~\cite{Eden:2016xvg,Fleury:2016ykk} functions, and even non-planar corrections~\cite{Eden:2017ozn,Bargheer:2017nne}. In $\AdS_5\times\S^5$ L\"uscher-like wrapping effects make such computations very hard, while we have seen in Eq.~\eqref{eq:wrapping} that those cancel here, at least for two-point functions. This, together with the wealth of data available might make NS-NS background an ideal playground for the hexagon bootstrap program~\cite{Basso:2015zoa,Eden:2016xvg, Fleury:2016ykk,Eden:2017ozn,Bargheer:2017nne}.

\begin{acknowledgments}
\paragraph{Acknowledgments.}
We are grateful to B.~Hoare, O.~Ohlsson Sax, B.~Stefa\'nski, G.~Tartaglino-Mazzucchelli, R.~Tateo and A.~Tseytlin for comments on the manuscript.
We thank A.~Dei, K.~Ferreira, B.~Hoare, S.~Komatsu, B.~Oblak, G.~Tartaglino-Mazzucchelli, L.~Wulff, K.~Zarembo for useful related discussions. AS is grateful to R.~Borsato, O.~Ohlsson~Sax, B.~Stefa{\'n}ski and A.~Torrielli, as well as to A.~Dei and M.~Gaberdiel for pleasant and fruitful collaboration on related topics.
The work of MB is supported by the European Union's Horizon 2020 research and innovation programme under the
Marie Sk\l odowska-Curie grant agreement no.~665501 with the Research Foundation Flanders (FWO). MB is an FWO [PEGASUS]$^2$ Marie
Sk\l odowska-Curie Fellow.
AS acknowledges support by the ETH ``Career Seed Grant'' n. 0-20313-17, as well as partial support from the NCCR SwissMAP, funded by the Swiss National Science Foundation.
\end{acknowledgments}

\end{document}